\documentclass[12pt]{article}
\usepackage{amssymb}

\def\R{\mbox{$\mathbb R$}}
\newcommand{\rx}{{\rm x}}
\newcommand{\rp}{{\rm p}}
\newcommand{\rhh}{{\rm H}}
\newcommand{\rh}{{\rm h}}
\newcommand{\rvv}{{\rm V}}
\newcommand{\rr}{{\rm r}}
\newcommand{\ri}{{\rm i}}
\newcommand{\rmm}{{\rm M}}
\newcommand{\rxx}{{\rm X}}
\newcommand{\rpp}{{\rm P}}
\newcommand{\rc}{{\rm c}}

\title{
Pseudo-Hermitian versus Hermitian position-dependent-mass Hamiltonians in a perturbative
framework}
\author{B Bagchi$^a$, C Quesne$^{b}$, R Roychoudhury$^c$\\ 
{\small
$^a$ Department of Applied Mathematics, University of Calcutta,} \\ {\small 92 Acharya
Prafulla Chandra Road, Kolkata 700 009, India}\\ 
{\small $^b$ Physique Nucl\'eaire
Th\'eorique et Physique Math\'ematique,  Universit\'e Libre de Bruxelles,} \\ 
{\small Campus de la Plaine CP229, Boulevard~du Triomphe, B-1050
Brussels, Belgium}\\
{\small $^c$ Physics and Applied Mathematics Unit, Indian Statistical Institute, Kolkata
700 035, India}\\ {\small E-mail: bbagchi123@rediffmail.com, cquesne@ulb.ac.be,
raj@isical.ac.in}}
\date{ }
\begin{document}
\baselineskip=22pt plus 1pt minus 1pt
\maketitle

\begin{abstract} 
We formulate a systematic algorithm for constructing a whole class of Hermitian
position-dependent-mass Hamiltonians which, to lowest order of perturbation theory, allow a
description in terms of $\cal PT$-symmetric Hamiltonians. The method is applied to the
Hermitian analogue of the $\cal PT$-symmetric cubic anharmonic oscillator. A new example is
provided by a Hamiltonian (approximately) equivalent to a $\cal PT$-symmetric extension of
the one-parameter trigonometric P\"oschl-Teller potential.  
\end{abstract}

\noindent
Keywords: Schr\"odinger equation, pseudo-Hermiticity, $PT$-symmetry, position-dependent
mass

\noindent
PACS No.: 03.65.-w
%
%
\newpage
Pseudo-Hermitian Hamiltonians  and their subclass of $\cal PT$-symmetric ones have aroused a
great deal of interest since it was observed that some of them may have a real, positive
spectrum~\cite{bender98}. Pseudo-Hermiticity of $H$ with respect to a positive-definite
(Hermitian and invertible) operator $\eta_+$, i.e.,
\begin{equation}
  H^{\dagger} = \eta_+ H \eta_+^{-1}  \label{eq:pseudo}
\end{equation}
has been identified  as one of the necessary and sufficient conditions for this situation to
occur~\cite{mosta02}. Any Hamiltonian endowed with such a property is then equivalent to a
Hermitian one
\begin{equation}
  h = \rho H \rho^{-1}  \label{eq:h}
\end{equation}
where the similarity transformation is implemented by $\rho = \sqrt{\eta_+}$. Further, to any
observable $o$ and to any wavefunction $\psi(x) = \langle x | \psi\rangle$ in the Hermitian
theory described by $h$, one can associate an operator $O = \rho^{-1} o \rho$ and a
wavefunction $\Psi(x) = \langle x | \rho \psi\rangle$ in the (physical) pseudo-Hermitian
theory, respectively.\par
%
%
Recently Jones~\cite{jones} and, independently, Mostafazadeh~\cite{mosta05}
constructed the Hermitian analogue $h$, as well as the pseudo-Hermitian position and
momentum operators $X = \rho^{-1} x \rho$, $P = \rho^{-1} p \rho$, for the $\cal
PT$-symmetric cubic anharmonic oscillator $H = \frac{1}{2}(p^2 + x^2) + {\rm i} \epsilon
x^3$ (with $\epsilon \in \R$). The latter, which has been shown both
numerically~\cite{bender98} and mathematically~\cite{dorey} to have a real, positive and
discrete spectrum, can only be treated in perturbation theory~\cite{bender03}. A very
interesting outcome of~\cite{jones} and~\cite{mosta05} is that to lowest order such a system
describes an ordinary quartic anharmonic oscillator with real and positive coupling constants
but a position-dependent mass (PDM). As revealed by a more recent study of Bender {\sl et
al\/}~\cite{bender05}, this Hermitian PDM theory is however difficult to work out because it
leads to divergent Feynman graphs, which must be regulated to obtain the correct answer,
whereas the corresponding non-Hermitian $\cal PT$-symmetric theory is completely free from
such difficulties.\par
%
%
At this stage, it is worth mentioning that Hermitian PDM Hamiltonians are attracting a lot of
attention due to their relevance in describing the physics of many microstructures of current
interest, such as compositionally graded crystals (see \cite{bagchi04} and references quoted
therein). Several classes of physically-interesting solvable non-Hermitian potentials have also
been generated~\cite{koc, roy, gonul} in a PDM background by employing various techniques,
such as the point canonical transformations or Lie algebraic methods, or using ideas from
supersymmetric quantum mechanics. In particular, constructions of $\cal PT$-symmetric
potentials have been carried out for different choices of mass functions. These include the
$\cal PT$-symmetric Scarf potential~\cite{koc} and the $\cal PT$-symmetric oscillator
model~\cite{roy}. Even the PDM version of the complex Morse potential~\cite{bagchi00},
which is known to be pseudo-Hermitian~\cite{ahmed}, has been obtained~\cite{roy}.\par
%
%
In view of all these considerations, it may prove interesting to see under which conditions a
Hermitian PDM Hamiltonian may be approximately equivalent to a non-Hermitian
$\cal PT$-symmetric one, which, according to the experience gained in~\cite{bender05}, would
presumably be easier to handle. In the spirit of~\cite{jones} and~\cite{mosta05}, this is
tantamount to determining those $\cal PT$-symmetric Hamiltonians, 
\begin{equation}
  H = H_0 + \varepsilon H_1 \qquad H_0 = \frac{p^2}{2m_0} + V^{(\rr)}(x) \qquad H_1 =
  {\ri} V^{(\ri)}(x)  \label{eq:H} 
\end{equation}
with $\varepsilon \in \R$, $V^{(\rr)}(x) = V^{(\rr)}(-x) \in \R$, $V^{(\ri)}(x) = -
V^{(\ri)}(-x) \in \R$ and configuration space \R\ (or a subset of it), that have a Hermitian
counterpart
\begin{equation}
  h = H_0 + \varepsilon^2 h^{(2)} + \varepsilon^4 h^{(4)} + \cdots  \label{eq:hbis}
\end{equation}
which to lowest order in $\varepsilon$ reduces to some PDM Hamiltonian, i.e.,
\begin{equation}
  H_0 + \varepsilon^2 h^{(2)} = p \frac{1}{2m(x)} p + V_{\rm eff}(x)  \label{eq:PDM-H}
\end{equation}
with $1/m(x) = (1/m_0) [1 + \varepsilon^2 M^{(2)}(x)]$, $V_{\rm eff}(x) = V^{(\rr)}(x)
+ \varepsilon^2 V^{(2)}_{\rm eff}(x)$ and $M^{(2)}(x)$, $V^{(2)}_{\rm eff}(x) \in
\R$. It should be noted that the right-hand side of (\ref{eq:hbis}) only contains even powers of
$\varepsilon$ because the coefficients of odd powers have been shown to vanish~\cite{jones,
mosta05}, while the right-hand side of (\ref{eq:PDM-H}) is the most general expression of
Hermitian PDM Hamiltonians~\cite{bagchi04}. The latter is written in terms of an effective
potential $V_{\rm eff}(x)$ including some mass terms depending on two ambiguity
parameters, which take the noncommutativity of the momentum and PDM operators into
account~\cite{vonroos}.\par
%
%
It proves convenient to introduce dimensionless quantities defined by
\begin{eqnarray}
  && \rx = \ell^{-1} x \qquad \rp = \ell \hbar^{-1} p \nonumber \\
  && \rhh = \nu^{-1} H = \rhh_0 + \varepsilon \rhh_1 \qquad \rhh_0 = \frac{1}{2} \rp^2 +
      \rvv^{(\rr)}(\rx) \qquad \rhh_1(\rx) = \ri \rvv^{(\ri)}(\rx) \nonumber \\
  && \rh = \nu^{-1} h = \rhh_0 + \varepsilon^2 \rh^{(2)} = \frac{1}{2} \rp [1 +
      \varepsilon^2 \rmm^{(2)}(\rx)] \rp + \rvv^{(\rr)}(\rx) + \varepsilon^2 \rvv^{(2)}_{\rm
      eff}(\rx)
\end{eqnarray}
in terms of some length and energy scales, $\ell$ and $\nu = \hbar^2/(m_0 \ell^2)$. Note
that in (\ref{eq:H}), (\ref{eq:hbis}) and (\ref{eq:PDM-H}), $\varepsilon$ is also dimensionless,
as well as $M^{(2)}(x)$.\par
%
%
In \cite{jones} and \cite{mosta05} (see also \cite{bender03}), it has been shown that for
the positive-definite metric operator $\eta_+$, one may take
\begin{equation}
  \eta_+ = e^{- Q(\rx,\rp)} \qquad Q(\rx,\rp) = \varepsilon Q_1(\rx,\rp) + \varepsilon^3
  Q_3(\rx,\rp) + \cdots
\end{equation}
where every $Q_j(\rx,\rp)$, $j=1$, 3,~\ldots, is such that $Q_j(\rx,\rp) =
Q_j^{\dagger}(\rx,\rp) = Q_j(-\rx,\rp) = - Q_j(\rx,-\rp)$. Then to lowest order in
$\varepsilon$, equations (\ref{eq:pseudo}) and (\ref{eq:h}) lead to the two conditions
\begin{equation}
  [\rhh_0, Q_1] = - 2\rhh_1 \qquad \frac{1}{4} [\rhh_1, Q_1] = \rh^{(2)}
\end{equation}
which in the case of (\ref{eq:H}) and (\ref{eq:PDM-H}) amount to
\begin{eqnarray}
  && \left[\frac{1}{2}\rp^2 + \rvv^{(\rr)}(\rx), Q_1\right] = - 2{\ri} \rvv^{(\ri)}(\rx) 
      \label{eq:C1}  \\
  && \frac{\ri}{4} [\rvv^{(\ri)}(\rx), Q_1] = \frac{1}{2} \rp \rmm^{(2)}(\rx) \rp +
      \rvv^{(2)}_{\rm eff}(\rx).  \label{eq:C2}
\end{eqnarray}
\par
%
%
{}For $Q_1$, let us choose a general ansatz somewhat different from those previously
considered:
\begin{equation}
  Q_1 = \sum_{k=0}^{\infty} \{R_k(\rx), \rp^{2k+1}\} \qquad R_k(\rx) = R_k(-\rx).
  \label{eq:Q1}
\end{equation}
By expressing $\rp$ as $- {\ri} d/d\rx$ and using the commutation relation
\begin{equation}
  \left[\frac{d^k}{d\rx^k}, f(\rx)\right] = \sum_{l=0}^{k-1} 
  \left(\begin{array}{c}
          k \\
          l
       \end{array}\right) 
  \frac{d^{k-l}f(\rx)}{d\rx^{k-l}} \frac{d^l}{d\rx^l}  \label{eq:com}
\end{equation}
$Q_1$ can be written in normal form, i.e., with all functions of $\rx$ on the left of the
differential operators, as
\begin{equation}
  Q_1 = - {\ri} \sum_{k=0}^{\infty} S_k(\rx) \frac{d^k}{d\rx^k}  \label{eq:Q1bis}
\end{equation}
where
\begin{eqnarray}
  && S_{2k} = \sum_{l=k}^{\infty} (-1)^l
       \left(\begin{array}{c}
          2l+1 \\
          2k
       \end{array}\right) 
       \frac{d^{2l-2k+1} R_l}{d\rx^{2l-2k+1}} \nonumber \\ 
  && S_{2k+1} = \sum_{l=k}^{\infty} (1 + \delta_{l,k}) (-1)^l 
        \left(\begin{array}{c}
        2l+1 \\
        2k+1
     \end{array}\right)
     \frac{d^{2l-2k} R_l}{d\rx^{2l-2k}}  \label{eq:S}
\end{eqnarray}
for $k=0$, 1, 2,~\ldots.\par
%
%
On inserting (\ref{eq:Q1bis}) in (\ref{eq:C1}) and (\ref{eq:C2}) and employing (\ref{eq:com})
again, we find after some straightforward calculations that equation (\ref{eq:C1}) is
equivalent to the conditions
\begin{eqnarray}
  && \frac{1}{2} \frac{d^2 S_0}{d\rx^2} + \sum_{l=1}^{\infty} S_l \frac{d^l \rvv^{(\rr)}}
        {d\rx^l} = - 2 \rvv^{(\ri)}  \label{eq:C11} \\
  && \frac{dS_{k-1}}{d\rx} + \frac{1}{2} \frac{d^2 S_k}{d\rx^2} + \sum_{l=k+1}^{\infty} 
        \left(\begin{array}{c}
          l \\
          k
        \end{array}\right)
        S_l \frac{d^{l-k}\rvv^{(\rr)}}{d\rx^{l-k}} = 0 \qquad k=1, 2, \ldots  \label{eq:C12}
\end{eqnarray}
while equation (\ref{eq:C2}) leads to 
\begin{eqnarray}
  && \sum_{l=1}^{\infty} S_l \frac{d^l \rvv^{(\ri)}}{d\rx^l} = - 4 \rvv^{(2)}_{\rm eff}
        \label{eq:C21} \\
  && \sum_{l=2}^{\infty}
        \left(\begin{array}{c}
          l \\
          1
        \end{array}\right)
        S_l \frac{d^{l-1} \rvv^{(\ri)}}{d\rx^{l-1}} = 2 \frac{d \rmm^{(2)}}{d\rx}  \label{eq:C22}
\\
  && \sum_{l=3}^{\infty}
        \left(\begin{array}{c}
          l \\
          2
        \end{array}\right)
        S_l \frac{d^{l-2} \rvv^{(\ri)}}{d\rx^{l-2}} = 2 \rmm^{(2)}  \label{eq:C23} \\
  && \sum_{l=k+1}^{\infty}
        \left(\begin{array}{c}
          l \\
          k
        \end{array}\right)
        S_l \frac{d^{l-k} \rvv^{(\ri)}}{d\rx^{l-k}} = 0 \qquad k=3, 4, \ldots.  \label{eq:C24}    
\end{eqnarray}
\par
%
%
To be able to solve the general equations (\ref{eq:C11})--(\ref{eq:C24}), it is appropriate to
make some simplifying assumption. Inspired by the example of the $\cal PT$-symmetric cubic
anharmonic oscillator considered in~\cite{jones, mosta05}, where $Q_1$ only contains linear
and cubic powers of $\rp$, let us assume that $R_k(\rx) = 0$, $k=2$, 3,~\ldots, in equation
(\ref{eq:Q1}). It then follows from (\ref{eq:S}) that only the first four functions $S_k$ in the
expansion (\ref{eq:Q1bis}) may be nonvanishing and that they are given in terms of $R_0$,
$R_1$, and their derivatives by $S_0 = R'_0 - R'''_1$, $S_1 = 2 R_0 - 3 R''_1$, $S_2 = - 3
R'_1$ and $S_3 = - 2 R_1$.\par
%
%
Let us first solve equations (\ref{eq:C11}) and (\ref{eq:C12}). In the latter, $k$ is now
restricted to $k \le 4$. For $k=4$, we obtain that $S_3$ must be a constant, this implying
that
\begin{equation}
  R_1(\rx) = c_1.
\end{equation}
Hence the remaining nonvanishing $S_k$'s are
\begin{equation}
  S_0 = R'_0 \qquad S_1 = 2 R_0 \qquad S_3 = - 2 c_1.  \label{eq:Sbis}
\end{equation}
From equation (\ref{eq:C12}) with $k=2$, we get
\begin{equation}
  R_0(\rx) = 3 c_1 \rvv^{(\rr)}(\rx) + c_0  \label{eq:R_0}
\end{equation}
where $c_0$ is another integration constant, while the equations with $k=1$ or $k=3$ are
automatically satisfied. Equation (\ref{eq:C11}) then provides us with a condition on
$\rvv^{(\ri)}$,
\begin{equation}
  \rvv^{(\ri)}(\rx) = \frac{1}{4} c_1 \rvv^{(\rr)\prime\prime\prime}(\rx) - [3 c_1
  \rvv^{(\rr)}(\rx) + c_0] \rvv^{(\rr)\prime}(\rx).  \label{eq:Vi}
\end{equation}
\par
%
%
Let us next turn ourselves to equations (\ref{eq:C21})--(\ref{eq:C24}). It is easy to see that
only equations (\ref{eq:C21}) and (\ref{eq:C23}) impose some new conditions, namely
\begin{equation}
  \rmm^{(2)}(\rx) = - 3 c_1 \rvv^{(\ri)\prime}(\rx) \qquad \rvv^{(2)}_{\rm eff}(\rx) =
  \frac{1}{2} \{- [3 c_1 \rvv^{(\rr)}(\rx) + c_0] \rvv^{(\ri)\prime}(\rx) + c_1 \rvv^{(\ri)
  \prime\prime\prime}(\rx)\}
  \label{eq:PDM}
\end{equation}
where $\rvv^{(\ri)}(\rx)$ must be expressed in terms of $\rvv^{(\rr)}(\rx)$ through equation
(\ref{eq:Vi}). This completes the solution of equations (\ref{eq:C1}) and (\ref{eq:C2}).\par
%
%
It is then straightforward to go back to $x$, $p$ and unscaled operators. This leads to the
conclusion that there exists a whole class of Hermitian PDM Hamiltonians, which to lowest
order of perturbation theory allow an equivalent $\cal PT$-symmetric description and might
therefore be easier to deal with than generic ones. The various members of the class are
distinguished by the choice of the zeroth-order part $V^{(\rr)}(x)$ of the effective potential
$V^{(2)}_{\rm eff}(x)$ and that of two integration constants $c_0$, $c_1$. The
lowest-order corrections to the mass term $M^{(2)}(x)$ and to the effective potential in the
PDM equation, as well as the imaginary part $V^{(\ri)}(x)$ of the corresponding $\cal
PT$-symmetric potential, are indeed entirely fixed by such a choice.\par
%
%
The classical Hamiltonians $H_\rc(x_\rc, p_\rc)$ corresponding to the members of this class
can be obtained by replacing $x$ and $p$ in $h$ by the classical variables $x_\rc$ and
$p_\rc$ and evaluating the resulting expressions in the limit $\hbar \to 0$ (assuming this
limit exists), i.e., $H_\rc(x_\rc, p_\rc) = \lim_{\hbar\to 0} h(x_\rc, p_\rc)$.\par
%
%
The $\eta_+$-pseudo-Hermitian position and momentum operators $X$ and $P$, as well as
the physical wavefunctions $\Psi(x)$, can be calculated in the same way as $h$. To second
order in $\varepsilon$, the pseudo-Hermitian operators are given by
\begin{equation}
  O = o - \frac{1}{2} \varepsilon [o, Q_1] + \frac{1}{8} \varepsilon^2 [[o, Q_1], Q_1] \qquad
  o = x \mbox{\ or\ } p.  \label{eq:O}
\end{equation}
For the dimensionless operators, we find
\begin{eqnarray}
  && [\rx, Q_1] = {\rm i} \sum_{k=0}^{\infty} (k+1) S_{k+1} \frac{d^k}{d\rx^k} \qquad
       [[\rx, Q_1], Q_1] = \sum_{k=0}^{\infty} T_k \frac{d^k}{d\rx^k} \nonumber \\
  && [\rp, Q_1] = - \sum_{k=0}^{\infty} \frac{dS_k}{d\rx} \frac{d^k}{d\rx^k} \qquad
       [[\rp, Q_1], Q_1] = {\rm i} \sum_{k=0}^{\infty} U_k \frac{d^k}{d\rx^k} 
\end{eqnarray}
where $T_k$ and $U_k$ are defined by
\begin{equation}
  T_k = \sum_{l=0}^k \sum_{m=k-l+1}^{\infty} T_k^{(l,m)} \qquad U_k = \sum_{l=0}^k
  \sum_{m=k-l+1}^{\infty} U_k^{(l,m)} 
\end{equation}
with
\begin{eqnarray}
  && T_k^{(l,m)} = 
          \left(\begin{array}{c}
          m\\
          k-l
        \end{array}\right)
        \left[(m+1) S_{m+1} \frac{d^{l+m-k} S_l}{d\rx^{l+m-k}} - (l+1) S_m \frac{d^{l+m-k}
        S_{l+1}}{d\rx^{l+m-k}}\right]  \nonumber \\
  && U_k^{(l,m)} = 
          \left(\begin{array}{c}
          m\\
          k-l
        \end{array}\right)
        \left[\frac{dS_m}{d\rx} \frac{d^{l+m-k} S_l}{d\rx^{l+m-k}} - S_m \frac{d^{l+m-k+1}
        S_l}{d\rx^{l+m-k+1}}\right].  \label{eq:T-U}  
\end{eqnarray}
Similarly, the physical wavefunctions can be expressed as
\begin{equation}
  \Psi(\rx) = \psi(x) - \frac{\varepsilon}{2} \langle\rx | Q_1 |\psi\rangle + \frac{\varepsilon^2}
  {8} \langle\rx | Q_1^2 |\psi\rangle
\end{equation}
where $Q_1$ is given by (\ref{eq:Q1bis}) and
\begin{equation}
  Q_1^2 = - \sum_{k=0}^{\infty} W_k(\rx) \frac{d^k}{d\rx^k}  
\end{equation}
with
\begin{equation}
  W_k = \sum_{l=0}^k \sum_{m=k-l}^{\infty} W_k^{(l,m)} \qquad W_k^{(l,m)} = 
  \left(\begin{array}{c}
        m \\
        k-l
     \end{array}\right)
  S_m \frac{d^{l+m-k}S_l}{d\rx^{l+m-k}}.  \label{eq:W}
\end{equation}
\par
%
%
With the simplifying assumption (\ref{eq:Sbis}) and taking equations (\ref{eq:R_0}) and
(\ref{eq:O})--(\ref{eq:W}) into account, we obtain
\begin{eqnarray}
  \rxx & = & \rx - {\ri} \varepsilon (3 c_1 \rvv^{(\rr)} + c_0 + 3 c_1 \rp^2) +
       \frac{3}{4} \varepsilon^2 c_1 [- c_1 (6 \rvv^{(\rr)} \rvv^{(\rr)\prime} + 
       \rvv^{(\rr)\prime\prime\prime}) - 2 c_0 \rvv^{(\rr)\prime}  \nonumber \\
  && \mbox{} - 6 {\ri} c_1 \rvv^{(\rr)\prime\prime} \rp + 6 c_1 \rvv^{(\rr)\prime} \rp^2]
       \label{eq:X}
\end{eqnarray}
\begin{eqnarray}
  \rpp & = & \rp + \frac{3}{2} \varepsilon c_1 (\rvv^{(\rr)\prime\prime} + 2 {\ri} 
        \rvv^{(\rr)\prime} \rp) + \frac{3}{4} {\ri} \varepsilon^2 c_1 \{c_1 (3 \rvv^{(\rr)
        \prime} \rvv^{(\rr)\prime\prime} - 3 \rvv^{(\rr)} \rvv^{(\rr)\prime\prime\prime} + 
        \rvv^{(\rr)\prime\prime\prime\prime\prime})  \nonumber \\
  && \mbox{} - c_0 V^{(\rm r)\prime\prime\prime} + {\rm i} [c_1 (6 V^{(\rm r)\prime2}
        - 6 \rvv^{(\rr)} \rvv^{(\rr)\prime\prime} + 5 \rvv^{(\rr)\prime\prime\prime\prime}) -
        2 c_0 \rvv^{(\rr)\prime\prime}] \rp - 9 c_1 \rvv^{(\rr)\prime\prime\prime} \rp^2
        \nonumber \\
  && \mbox{} - 6 {\ri} c_1 \rvv^{(\rr)\prime\prime} \rp^3\}  \label{eq:P} 
\end{eqnarray}
and
\begin{eqnarray}
  \Psi(\rx) & = & \psi(x) + \frac{1}{2} \ri \varepsilon \left[3 c_1 \rvv^{(\rr)\prime} + 2
      (3 c_1 \rvv^{(\rr)} + c_0) \frac{d}{d\rx} - 2 c_1 \frac{d^3}{d\rx^3}\right]
      \nonumber \\ 
  && \mbox{} -\frac{\varepsilon^2}{8}\Biggl\{3 c_1 [c_1(3 \rvv^{(\rr)\prime2} + 6
      \rvv^{(\rr)} \rvv^{(\rr)\prime\prime} - 2 \rvv^{(\rr)\prime\prime\prime\prime}) + 2
      c_0 \rvv^{(\rr)\prime\prime}] \nonumber \\
  && \mbox{} + 6 c_1 [c_1(12 \rvv^{(\rr)} \rvv^{(r)\prime} - 5 \rvv^{(\rr)\prime\prime
      \prime}) + 4 c_0 \rvv^{(r)\prime}] \frac{d}{d\rx} + 2 [9 c_1^2 (2 \rvv^{(\rr)2} - 3
      \rvv^{(\rr)\prime\prime}) \nonumber \\
  && \mbox{} + 12 c_0 c_1 \rvv^{(\rr)} + 2 c_0^2] \frac{d^2}{d\rx^2} - 48 c_1^2
      \rvv^{(\rr)\prime} \frac{d^3}{d\rx^3} - 8c_1 (3c_1 \rvv^{(\rr)} + c_0) 
      \frac{d^4}{d\rx^4}  \nonumber \\
  && \mbox{} + 4 c_1^2 \frac{d^6}{d\rx^6}\Biggr\}.  \label{eq:Psi}
\end{eqnarray}
\par
%
%
It is easy to check that, as expected, the Hermitian PDM quartic anharmonic oscillator
of~\cite{jones, mosta05} belongs to the class of Hermitian PDM Hamiltonians with an
approximate $\cal PT$-symmetric counterpart. On setting $\rvv^{(\rr)}(\rx) =
\frac{1}{2} {\cal M}^2 \rx^2$, $c_0 = 0$ and $c_1 = - 2/(3{\cal M}^4)$ in equation
(\ref{eq:Vi}), where the dimensionless quantitites are defined as in equations (17)--(20)
of~\cite{mosta05}, we indeed obtain $\rvv^{(\ri)}(\rx) = \rx^3$, so that $V^{(\ri)}(x) =
x^3$. Furthermore, from equations (\ref{eq:PDM}), (\ref{eq:X}) and (\ref{eq:P}), we obtain
$m(x) = m_0 [1 + 6 (\epsilon^2/\mu^4) x^2]^{-1}$, $V^{(2)}_{\rm eff} (x) = (3 m_0
\mu^2 x^4 - 4 \hbar^2)/(2m_0 \mu^4)$, $X = x + \ri (\epsilon/\mu^4)(\mu^2 x^2 +
2p^2/m_0) + (\epsilon^2/\mu^6)(- \mu^2 x^3 - 2 \ri \hbar p/m_0 + 2 x p^2/m_0)$ and $P
= p - \ri (\epsilon/\mu^2)(2xp - \ri \hbar) + (\epsilon^2/\mu^6)(2 p^3/m_0 - \mu^2 x^2 p
+ \ri \hbar \mu^2 x)$, which after some reordering agree with~\cite{jones, mosta05}, as does
the classical Hamiltonian. Similarly, equation (\ref{eq:Psi}) gives rise to equation (65)
of~\cite{mosta05}.\par
%
%
A new example is provided by selecting for $V^{(\rr)}(x)$ a one-parameter  trigonometric
P\"oschl-Teller potential~\cite{nieto}
\begin{equation}
  V^{(\rr)}(x) = V_0 \sec^2 kx \qquad V_0 = \frac{\hbar^2 k^2}{2m^2} \lambda
  (\lambda - 1) \qquad \lambda > 2  \label{eq:Vr-new}
\end{equation}
on the interval $- \frac{\pi}{2} \le x \le \frac{\pi}{2}$. On setting $\ell = k^{-1}$ and $\nu
= \hbar^2 k^2/m_0$ for the length and energy scales, respectively, we obtain the
dimensionless quantities $\rx = kx$, $\rp = p/(\hbar k)$ and $\rvv^{(\rr)}(\rx) =
\frac{1}{2} \lambda (\lambda-1) \sec^2 \rx$ with $\lambda (\lambda-1) = 2 V_0/\nu$.\par
%
%
The choice $c_0 = - c_1 = \frac{1}{3}$ in (\ref{eq:Vi}) leads to 
\begin{equation}
  \rvv^{(\ri)}(\rx) = \frac{1}{2} (\lambda+1)\lambda(\lambda-1)(\lambda-2) \sec^4 \rx 
  \tan \rx = \frac{2}{\nu^2} V_0 (V_0-\nu) \sec^4\rx \tan \rx.
\end{equation}
This means that the corresponding ${\cal PT}$-symmetric Hamiltonian may be written as
\begin{equation}
  H = \frac{p^2}{2m_0} + V_0 \sec^2 kx + \ri \epsilon \sec^4 kx \tan kx \label{eq:H-new}
\end{equation}
where $\epsilon$ has the dimension of an energy and is given in terms of the dimensionless
$\varepsilon$ by $\epsilon = 2\varepsilon V_0 (V_0-\nu)/\nu$.\par
%
%
To second order in $\epsilon$, such a non-Hermitian Hamiltonian is equivalent to a Hermitian
PDM one, given by equation (\ref{eq:PDM-H}), where
\begin{equation}
  m(x) = m_0 \left(1 + \frac{\epsilon^2}{2V_0 (V_0-\nu)} \sec^4 kx (5 \sec^2 kx -
  4)\right)^{-1}
\end{equation}
and 
\begin{equation}
  V_{\rm eff}(x) = V_0 \sec^2 kx + \frac{\epsilon^2}{4V_0 (V_0-\nu)} \sec^4 kx [5(V_0  -
  14\nu) \sec^4 kx - (4V_0 - 85\nu) \sec^2 kx - 20\nu].
\end{equation}
The corresponding $\eta_+$-pseudo-Hermitian position and momentum operators can be
expressed as
\begin{eqnarray}
  X & = & x - \ri \frac{\epsilon}{2kV_0(V_0-\nu)} \left(- V_0 \sec^2 kx + \frac{\nu}{3} - 
      \frac{p^2}{m_0}\right) - \frac{\epsilon^2}{4kV_0 (V_0-\nu)^2} \sec^2 kx \nonumber
      \\ 
  && \mbox{} \times \left\{[(V_0 + 2\nu) \sec^2 kx - \nu] \tan kx + \ri
      \sqrt{\frac{\nu}{m_0}} (3 \sec^2 kx - 2) p - \tan kx \frac{p^2}{m_0} \right\} \\
  P & = & p - \frac{\epsilon}{2(V_0-\nu)} \sec^2 kx [\sqrt{m_0\nu} (3 \sec^2 kx - 2) +
      2\ri \tan kx\; p] - \ri \frac{\epsilon^2}{4V_0 (V_0-\nu)^2} \sec^2 kx \nonumber \\
  && \mbox{} \times \Biggl\{\sqrt{m_0\nu} [3V_0 \sec^4 kx - 2\nu (30 \sec^4 kx - 19
      \sec^2 kx + 1)] + \ri [V_0 \sec^4 kx  \nonumber \\
  && \mbox{}  - \nu (50 \sec^4 kx - 49 \sec^2 kx + 6)] p + 6 \sqrt{\frac{\nu}{m_0}} (3
      \sec^2 kx - 1) \tan kx\; p^2 \nonumber \\ 
  && \mbox{}+ \frac{\ri}{m_0} (3 \sec^2 kx - 2 ) p^3\Biggr\}.     
\end{eqnarray}
Similar results can be found for physical wavefunctions. For lack of space, let us only
mention the result in dimensionless variable obtained for the function $\psi(\rx) =
\cos^{\lambda}(\rx)$ (corresponding to the ground state of the real potential
(\ref{eq:Vr-new})):  
\begin{eqnarray}
  \Psi(\rx) & = & \cos^{\lambda}(\rx) \Biggl\{1 + \frac{\ri}{6} \varepsilon (\lambda+1)
      \lambda(\lambda-1) (\sec^2 \rx + 2) \tan \rx - \frac{\varepsilon^2}{72} (\lambda+1)
      \lambda(\lambda-1) \nonumber \\
  && \mbox{} \times [(\lambda-4)(\lambda-2)(\lambda+15) \sec^6 \rx + 3(\lambda-2)
      (\lambda^2 - 4\lambda + 15) \sec^4 \rx \nonumber \\
  && \mbox{} - 4(\lambda+1)\lambda(\lambda-1)\Biggr\}.   
\end{eqnarray}
\par
%
%
In the classical limit, $\nu$ goes to zero. To get a nonvanishing limit for $V_0$, we must
therefore assume that $\lambda$ goes to infinity as $\hbar^{-1}$ (this implying, in particular,
that $\lambda$ becomes negligeably small compared with $\lambda^2$). To second order in
$\epsilon$, the classical Hamiltonian corresponding to (\ref{eq:H-new}) is obtained as
\begin{equation}
  H_\rc = \frac{p_\rc^2}{2m(x_\rc)} + V_0 \sec^2 kx_\rc + \frac{\epsilon^2}{4V_0}
  \sec^6 kx_\rc (5 \sec^2 kx_\rc - 4)  \label{eq:Hc-new}  
\end{equation}
where
\begin{equation}
  m_\rc(x_\rc) = m_0 \left(1 - \frac{\epsilon^2}{2V_0^2} \sec^4 kx_\rc (5 \sec^2 kx_\rc -
  4)\right) 
\end{equation}
while the classical $\eta_+$-pseudo-Hermitian variables $X_\rc$, $P_\rc$ are
\begin{eqnarray}
  X_\rc & = & x_\rc + \ri \frac{\epsilon}{2kV_0^2} \left(V_0 \sec^2 kx_\rc +
       \frac{p_\rc^2}{m_0}\right) - \frac{\epsilon^2}{4kV_0^3} \sec^2 kx_\rc
       \nonumber \\
  && \mbox{} \times \left(V_0 \sec^2 kx_\rc - \frac{p_\rc^2}{m_0}\right) \tan kx_\rc \\
  P_\rc & = & p_\rc - \ri \frac{\epsilon}{V_0} \sec^2 kx_\rc \tan kx_\rc\;  p_\rc
       +\frac{\epsilon^2}{4V_0^3} \sec^2 kx_\rc \nonumber \\  
  && \mbox{} \times \left[V_0 \sec^4 kx_\rc + (3 \sec^2 kx_\rc - 2)
       \frac{p_\rc^2}{m_0}\right] p_\rc.    
\end{eqnarray}
It is worth noting that in contrast with what happens for the $\cal PT$-symmetric cubic
anharmonic oscillator, the operators $X$ and $P$ involve $\hbar$ even after rewritting them
in a symmetrized form. As a consequence, the $\eta_+$-pseudo-Hermitian quantization of the
classical Hamiltonian (\ref{eq:Hc-new}) is far from trivial. This illustrates the importance of the
factor-ordering problem in pseudo-Hermitian quantum mechanics.\par
%
%
In conclusion, the generalization of the works in~\cite{jones} and~\cite{mosta05} that we have
proposed here contributes to exploring further the relationships between $\cal
PT$-symmetric and Hermitian PDM Hamiltonians started there and continued
in~\cite{bender05, bagchi05}. Moreover, it suggests the interest of performing detailed
calculations for some new $\cal PT$-symmetric systems, such as the one defined in
(\ref{eq:H-new}).\par
%
%
\section*{Acknowledgments}

BB and RR gratefully acknowledge the support of the National Fund for
Scientific Research (FNRS), Belgium, and the warm hospitality at PNTPM, Universit\'e Libre
de Bruxelles, where this work was initiated. CQ is a Research Director of the National Fund for
Scientific Research (FNRS), Belgium.\par
%
%

\end{document}